# Tuning of ferromagnetism through anion substitution in Ga-Mn-pnictide ferromagnetic semiconductors

Peter. R. Stone, and Oscar D. Dubon
*Department of Materials Science & Engineering, University of California, Berkeley, CA 94720 and Lawrence Berkeley National Laboratory, Berkeley, CA 94720*

Jeffrey W. Beeman, and Kin M. Yu
*Lawrence Berkeley National Laboratory, Berkeley, CA 94720*

ABSTRACT

We have synthesized $Ga_{1-x}Mn_xAs_{1-y}P_y$ and $Ga_{1-x}Mn_xP_{1-y}N_y$ by the combination of ion implantation and pulsed-laser melting. We find that the incorporation of isovalent impurities with smaller atomic radii leads to a realignment of the magnetic easy axis in $Ga_{1-x}Mn_xP_{1-y}N_y$/GaP and $Ga_{1-x}Mn_xAs_{1-y}P_y$/GaAs thin films from in-plane to out-of-plane. This tensile-strain-induced magnetic anisotropy is reminiscent of that observed in $Ga_{1-x}Mn_xAs$ grown on larger lattice constant (In,Ga)As buffer layers indicating that the role of strain in determining magnetic anisotropy is fundamental to III-Mn-V materials. In addition, we observe a decrease in the ferromagnetic Curie temperature in $Ga_{1-x}Mn_xAs_{1-y}P_y$ with increasing y from 0 to 0.028. Such a decrease may result from localization of holes as the P/As ratio on the Group V sublattice increases.

## 1. Introduction

The discovery that conventional III-V semiconductors such as GaAs exhibit ferromagnetism at relatively high temperatures when doped with a few atomic percent Mn has led to significant research exploring the origin of this effect **[1]**. Ferromagnetic exchange in these so-called ferromagnetic semiconductors is mediated by holes provided by substitutional Mn acceptors ($Mn_{Ga}$). While significant research has been devoted to the $Ga_{1-x}Mn_xAs$ system in recent years, the effect of anion substitution - i.e. altering the group V element of the semiconductor host - on ferromagnetic exchange remains unclear. One can identify two fundamental effects that occur upon changing the group V element in the semiconductor host to, for example, one with a smaller atomic radius. A wider gap semiconductor host, such as GaP, has band edges which are closer energetically to the Mn-derived 3*d* levels. Thus, there is increased mixing of Mn 3*d* and anion *p* states leading to an enhancement of *p-d* exchange. On the other hand, the Mn acceptor level lies much deeper in the forbidden gap of the wider gap semiconductor hosts. This leads to increasingly localized hole wavefunctions which are less effective at mediating exchange between the dilute $Mn_{Ga}$ spins. It is, therefore, of great fundamental importance to explore how the interplay of these two effects affects the ferromagnetic properties of samples throughout the Ga-Mn-pnictide series. To this extent, our group has previously demonstrated the successful synthesis of $Ga_{1-x}Mn_xP$ in which ferromagnetic exchange is mediated by localized carriers located in a detached impurity band **[2]**. In this work we explore the effect of partial isovalent anion substitution on the ferromagnetic Curie temperature ($T_C$), magnetic anisotropy, and electronic transport in $Ga_{1-x}Mn_xAs_{1-y}P_y$ and $Ga_{1-x}Mn_xP_{1-y}N_y$ for $y \leq 0.028$. These novel materials systems allow us to systematically study how the aforementioned properties change over a range of compositions between the more well studied end-point compounds, thus shedding further light on ferromagnetic exchange in $Ga_{1-x}Mn_x$-V ferromagnetic semiconductors.

## 2. Experimental Procedure

All samples were synthesized using the combination of ion implantation and pulsed laser melting (II-PLM), the details of which can be found elsewhere **[3, 4]**. $Ga_{1-x}Mn_xAs$ ($Ga_{1-x}Mn_xP$) was synthesized by implanting wafers of (001)-oriented SI-GaAs (unintentionally *n*-doped GaP) at 7° from normal with 50 kV $Mn^+$ to a dose of either 0.5 or 1.5 x $10^{16}$ $cm^{-2}$. Samples were irradiated in air with a single pulse from a KrF ($\lambda$=248 nm) excimer laser having FWHM of 18 ns at a fluence of 0.3 $J/cm^2$ for $Ga_{1-x}Mn_xAs$ and 0.44 $J/cm^2$ for $Ga_{1-x}Mn_xP$. Quaternary thin films were made by performing further implantation prior to the laser processing. For $Ga_{1-x}Mn_xAs_{1-y}P_y$ the $Mn^+$-implanted GaAs was co-implanted with 60 kV $P^+$ to a dose of 7.5 x $10^{15}$ $cm^{-2}$. $Ga_{1-x}Mn_xP_{1-y}N_y$ was made by co-implanting the $Mn^+$-implanted GaP with 33 kV $N^+$ to a dose of 3.0x$10^{15}$ $cm^{-2}$. GaAs-based films were etched for 20 minutes in concentrated HCl to remove surface oxide layers **[5]**. GaP-based films were etched for 24 hours in concentrated HCl to remove oxides as well as a highly-twinned surface layer **[3]**. Dopant concentrations and substitutional fractions were determined by the combination of SIMS and ion beam analysis. We define x the peak $Mn_{Ga}$ concentration. The parameter y is defined analogously for the peak substitutional concentrations on the anion sublattice. DC magnetization was measured using a standard SQUID magnetometer.

## 3. Results and Discussion

### *3.1 - Magnetic Anisotropy: Strain-Engineered Out-of-Plane Easy Axis*

$Ga_{1-x}Mn_xAs$ films grown on GaAs substrates are under compressive strain due to the larger Mn atoms substituting on Ga sites. In $Ga_{1-x}Mn_xAs_{1-y}P_y$ grown on GaAs the $P_{As}$ sites counterbalance this compression as illustrated by the X-ray rocking curves in Figure 1. The film with x=0.036 is characterized by the main GaAs substrate peak at 33.02 degrees with a broad low-angle shoulder corresponding to the $Ga_{1-x}Mn_xAs$ film. That we see a shoulder and not a peak is a result of the non-uniform dopant concentration as a function of depth characteristic of films produced by II-PLM at this Mn concentration. Addition of only 2.8 % $P_{As}$ on the group V sublattice leads to a marked change in the diffraction profile. The low-angle feature decreases in intensity by over an order of magnitude while a new, strong shoulder develops on the high angle (smaller lattice constant) side of the main GaAs peak. The manganese and phosphorus have slightly different concentration profiles. Therefore, there exists a region of the film under compressive strain even though the majority of the film is in tension, thus explaining the multiple non-substrate features present in the $Ga_{0.962}Mn_{0.038}As_{0.972}P_{0.028}$ rocking curves.

The change in the film's strain state has a significant effect on the magnetic anisotropy. This is illustrated in Figure 2, which shows the field-dependence of the magnetization for the same two samples that were shown in Figure 1. The hysteresis loops were measured with the applied magnetic field normal to the sample plane at a temperature of 5 K. It is clear from Figure 2 that the addition of a few atomic percent P to $Ga_{1-x}Mn_xAs$ results in a reorientation of the magnetic easy axis from in-plane to out-of-plane, which we attribute to the tensile strain in the $Ga_{1-x}Mn_xAs_{1-y}P_y$ film This is consistent with results obtained from low temperature molecular beam epitaxy (LT-MBE) grown materials **[6]**. Consequently, the out-of-plane easy axis observed in tensile-strained $Ga_{1-x}Mn_xAs$ is not due to growth mechanisms, but is, in fact, an intrinsic property of strain-state of the ferromagnetic semiconductor film.

While the magnetic anisotropy of the $Ga_{1-x}Mn_xAs$ system has been extensively studied, comparatively little is known about magnetic anisotropy in other $Ga_{1-x}Mn_x$-pnictides. Similar to the case in $Ga_{1-x}Mn_xAs$ the easy axis of $Ga_{1-x}Mn_xP$ grown under compressive strain lies near the in plane [110] direction, with the out-of-plane orientation being magnetically hardest **[7]**. The addition of nitrogen allows us to grow $Ga_{1-x}Mn_xP$-based films in tensile strain. The out-of-plane hysteresis loops for $Ga_{0.982}Mn_{0.018}P$ films grown with and without N are shown in Figure 3. As in the $Ga_{1-x}Mn_xAs_{1-y}P_y$ system, a clear reorientation of the easy axis is observed when the host anion is partially substituted by a smaller, isovalent species. The value of y in the $Ga_{0.982}Mn_{0.018}P_{1-y}N_y$ film could not be resolved using standard ion beam analysis. However, based on our knowledge of implant parameters and PLM regrowth we estimate the value of y corresponding to a 33 kV $3.0 \times 10^{15}$ $cm^{-2}$ $N^+$ implant dose to be 0.01. We verified that nitrogen was substitutionally incorporated into the film and affected the strain through X-ray diffraction. A prominent high angle feature develops in films with N that is not present in $Ga_{0.982}Mn_{0.018}P$ demonstrating that the film is in tensile strain (not shown). This suggests that rotation of the magnetic easy axis from in-plane to out-of-plane by means of tensile strain is fundamental to $Ga_{1-x}Mn_x$-pnictide ferromagnetic semiconductors.

*3.2 - The Effect of Partial Anion Substitution on $T_C$*

As mentioned in Section 1 there are two competing effects that occur when changing the host semiconductor in Mn-based ferromagnetic semiconductors. To date, experimental evidence suggests that among single phase, Ga-pnictide host materials, $Ga_{1-x}Mn_xAs$ lies at the apex where *p-d* hybridization and carrier delocalization are simultaneously optimized. However, the $T_C$ record of 173 K in $Ga_{1-x}Mn_xAs$ is still well below room temperature **[8]**. It has been proposed that a possible route to enhance $T_C$ for a given value of x is to use a $GaAs_{1-y}P_y$ ternary semiconductor host for the $Mn_{Ga}$ moments **[9]**. By adding a wider band gap component to the host it is believed that one can maintain the itinerancy of the mediating holes while enhancing *p-d* exchange due to the shorter Mn-anion bond length. Recent calculations have, in fact, predicted an enhancement of $T_C$ by a factor of 1.5 due to this effect **[10]**. However, experimentally we find that $Ga_{1-x}Mn_xAs_{1-y}P_y$ shows the opposite behavior. This is illustrated in Figure 4 where the magnetization as a function of temperature is plotted for $Ga_{0.964}Mn_{0.036}As$ and $Ga_{0.962}Mn_{0.038}As_{0.972}P_{0.028}$. The theromagnetic curves in Figure 4 were measured with the applied field of 50 Oe parallel to the film's easy axis, thus allowing for a more accurate comparison of $T_C$. A reduction of $T_C$ with y was also observed in the $Ga_{1-x}Mn_xP_{1-y}N_y$ system. Here, a sample with x = 0.018 had its $T_C$ of 20 K reduced to 14 K with the addition of ~1% $N_P$ defects.

Given that charge transport and exchange are intimately related in ferromagnetic semiconductors, we have measured the temperature dependence of the resistivity for the two samples shown in Figure 4. The sample with no phosphorous exhibits metallic transport as expected for $Ga_{1-x}Mn_xAs$ films with sufficiently high $Mn_{Ga}$ concentration. Markedly different transport behavior is observed for the film that contains phosphorus; the resistivity tends to infinity at low temperature, indicative of non-metallic transport. Thus, the addition of the isovalent species P to the $Ga_{1-x}Mn_xAs_{1-y}$ system leads to a metal-insulator transition (MIT) even as the Mn dopant concentration is held at a near constant value. At this point we are unsure as to the origin of this effect. Regardless of the underlying mechanism, it is clear that deviation from the pure GaAs semiconductor host towards GaP leads to the localization of holes.

## 4. Summary

We have investigated the dependence of the magnetic anisotropy and ferromagnetic Curie temperature on anion sublattice composition in $Ga_{1-x}Mn_x$-pnictide ferromagnetic semiconductors. We have determined that the relationship between tensile strain and perpendicular magnetic anisotropy is both processing and materials independent, and, thus, fundamental to $Ga_{1-x}Mn_x$-pnictide ferromagnetic semiconductors. The introduction of smaller, isovalent P impurities into the underlying GaAs host both diminishes $T_C$ and localizes holes. Hence, the alloying of III-Mn-V ferromagnetic semiconductor films with dilute concentrations of smaller radius anions may not be a plausible route for the enhancement of $T_C$.

## Acknowledgements

The authors thank R. Farshchi and S.K.Z. Tardif for experimental assistance. This work is supported by the Director, Office of Science, Office of Basic Energy Sciences, Division of Materials Sciences and Engineering, of the U.S. Department of Energy under Contract No. DE-AC02-05CH11231. PRS acknowledges support from an NDSEG fellowship.

FIGURE CAPTIONS

**Figure 1 -** X-ray rocking curves about the (004) reflection for $Ga_{0.964}Mn_{0.036}As$ (black dashes line) and $Ga_{0.962}Mn_{0.038}As_{0.972}P_{0.028}$ (solid grey line).

**Figure 2** - Magnetization as a function of magnetic field for $Ga_{0.964}Mn_{0.036}As$ (black) and $Ga_{0.962}Mn_{0.038}As_{0.972}P_{0.028}$ (grey) measured at 5 K with the field applied perpendicular to the plane of the sample.

**Figure 3 -** Magnetization as a function of magnetic field for $Ga_{0.982}Mn_{0.018}P$ with (grey) and without (black) nitrogen incorporated on the anion sublattice. Loops were measured at 5 K with the field applied perpendicular to the plane of the sample. The concentration of substitutional nitrogen is estimated to be ~1%.

**Figure 4 -** Magnetization vs. temperature curves for $Ga_{0.964}Mn_{0.036}As$ and $Ga_{0.962}Mn_{0.038}As_{0.972}P_{0.028}$ measured at an applied field of 50 Oe. The field was oriented parallel to the magnetic easy axis of each film to ensure the most accurate representation of $T_C$.

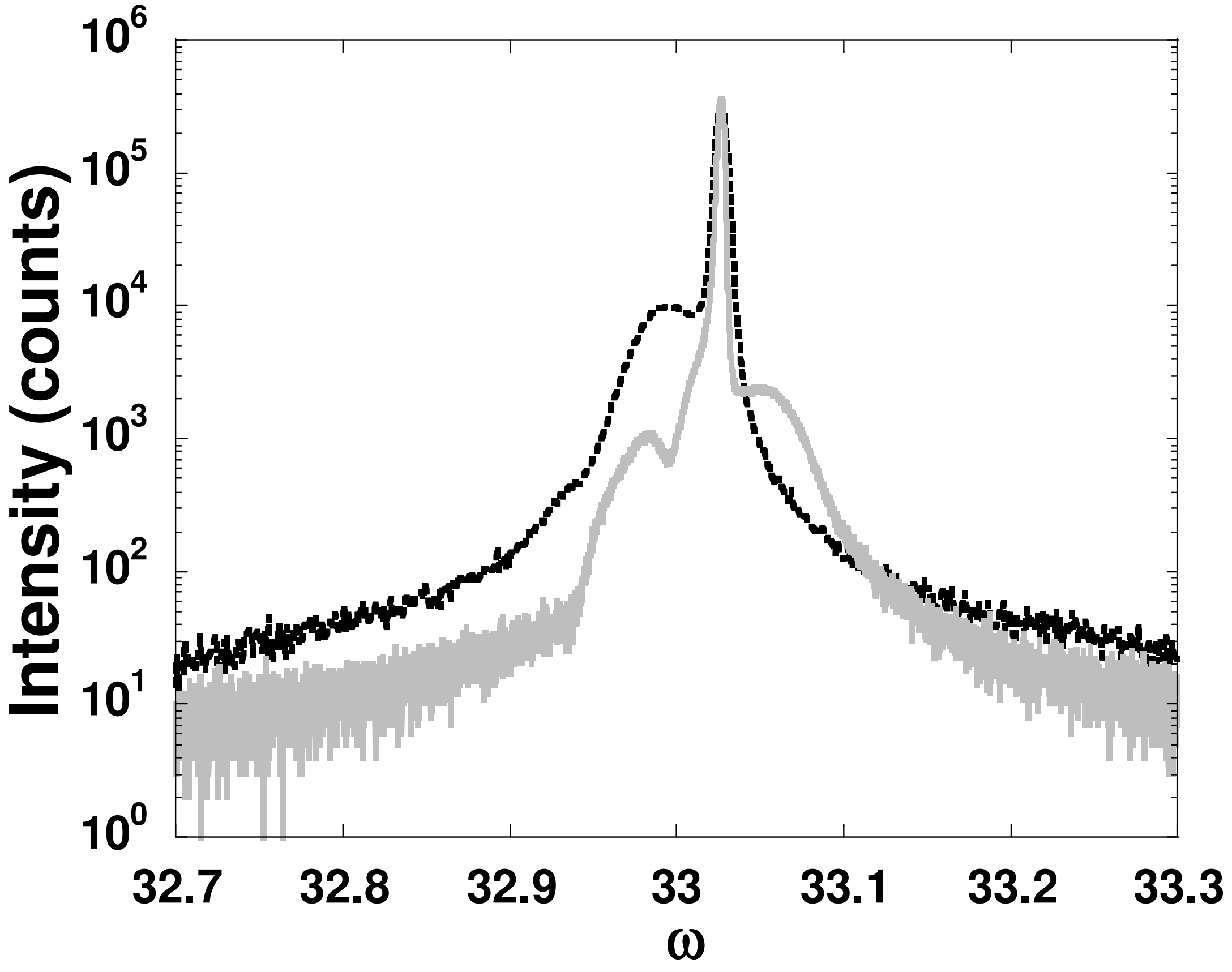

**Figure 1**

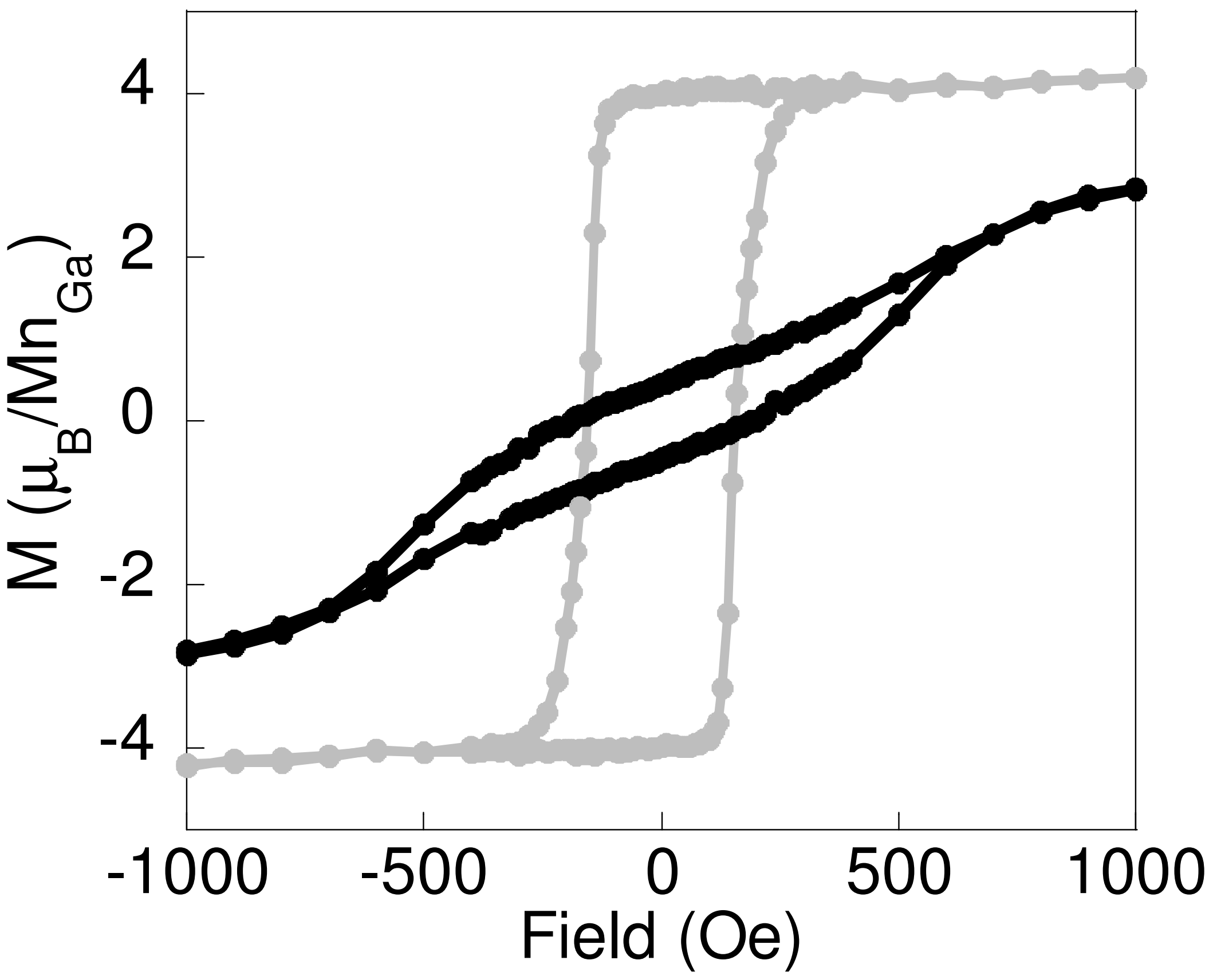

**Figure 2**

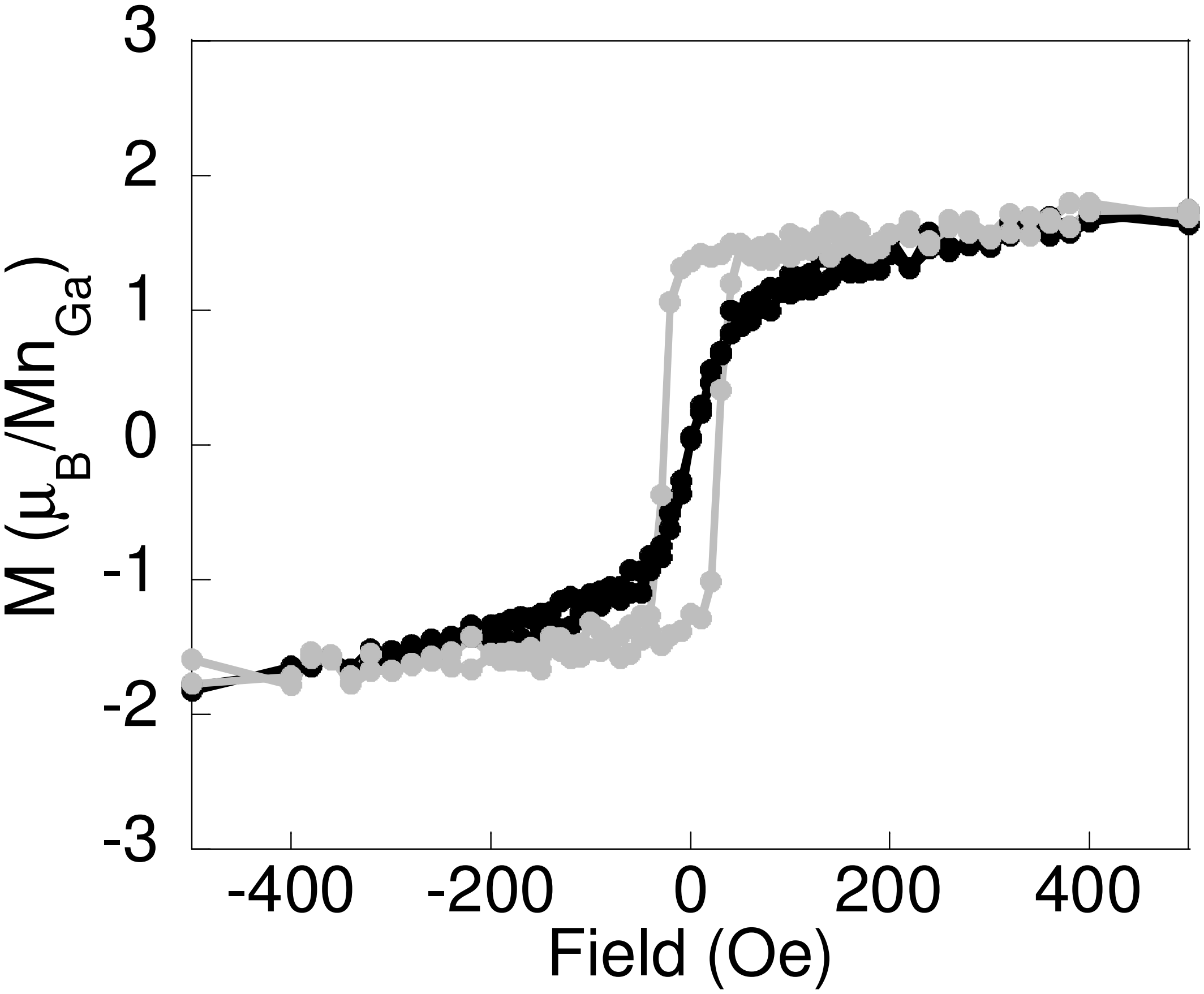

**Figure 3**

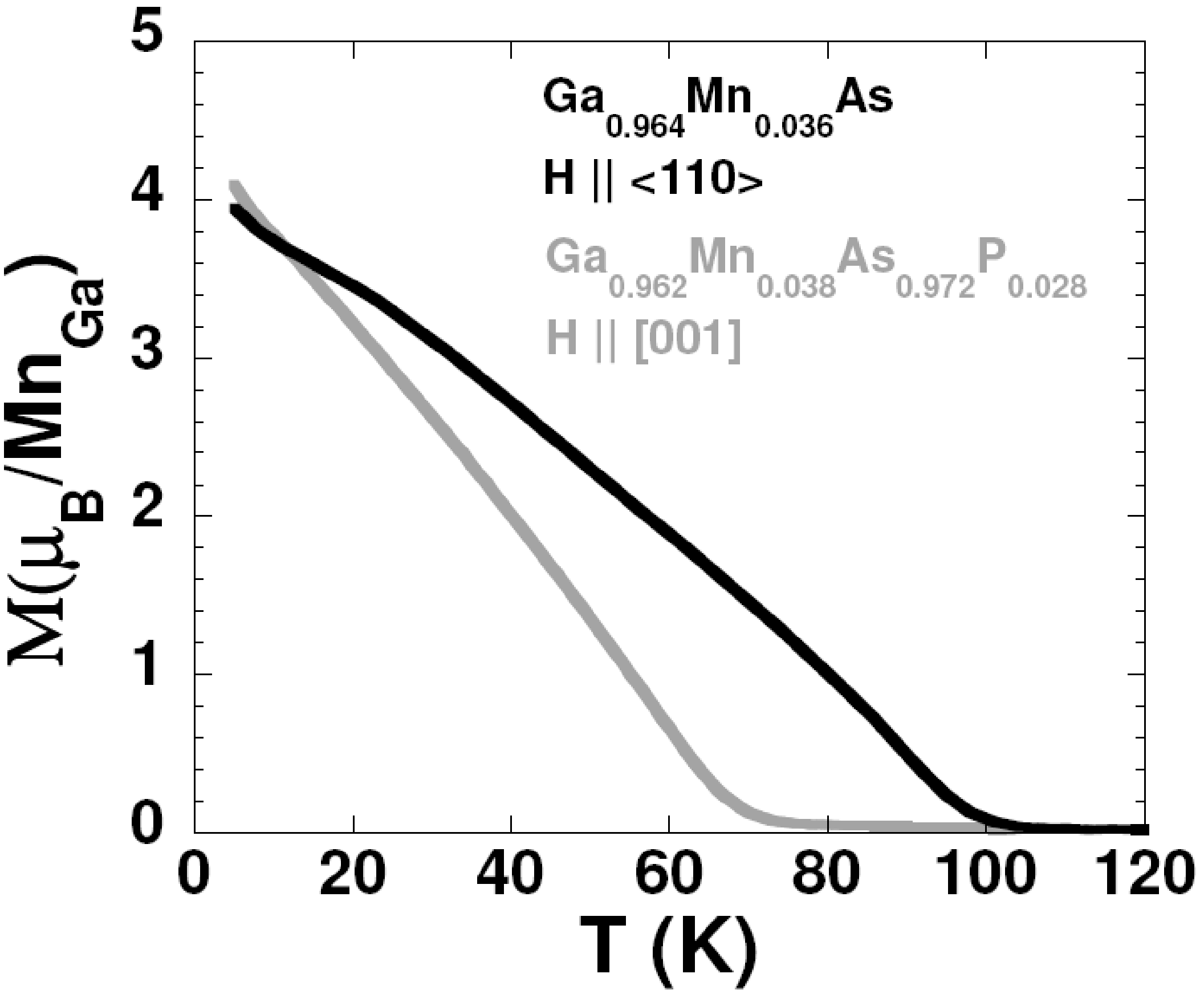

**Figure 4**